\documentclass[aps,twocolumn,prb]{revtex4}
 \usepackage{graphicx}
\usepackage{amsmath}
\usepackage{amssymb}
\newcommand\dd{{\operatorname{d}}}
\newcommand\sgn{{\operatorname{sgn}}}
\def\Eq#1{{Eq.~(\ref{#1})}}
\def\Ref#1{(\ref{#1})}
\newcommand\e{{\mathrm e}}
\newcommand\cum[1]{  {\Bigl< \!\! \Bigl< {#1} \Bigr>\!\!\Bigr>}}
\newcommand\vf{v_{_\text{F}}}
\newcommand\pf{p_{_\text{F}}}
\newcommand\ef{{\varepsilon} _{\text{\sc f}}}
\newcommand\zf{z_{_\text{F}}}
\newcommand\zfi[1]{{z_{_\text{F}}}_{#1}}
\newcommand\av[1]{\left<{#1}\right>}

\begin{document}

\title{Functional Integral Bosonization for Impurity in Luttinger Liquid}
\author{Alex Grishin, Igor V.\ Yurkevich and Igor V. Lerner}

 \affiliation{School of Physics and Astronomy, University of
Birmingham,
 Birmingham B15 2TT, United Kingdom}
\date{\today}

\begin{abstract}
We use a functional integral formalism developed earlier for the
pure Luttinger liquid (LL) to find an exact representation for the
electron Green  function of the LL in the presence of a single
backscattering impurity. This allows us to reproduce results (well
known from the bosonization techniques)  for the suppression of
the electron local density of states (LDoS) at the position of the
impurity and for the Friedel oscillations at finite temperature.
In addition, we have extracted from the exact representation an
analytic dependence of LDoS on the distance from the impurity and
shown how it crosses over to that for the pure LL.
\end{abstract}

\maketitle

\section{Introduction}
\setlength{\mathindent}{0cm} \setlength{\unitlength}{1cm}

One of the first exactly soluble models in the problem of strongly
correlated electrons was formulated in one dimension in the
seminal papers of Tomonaga \cite{Tom:50} and Luttinger
\cite{Lutt:63} and solved by Mattis and Lieb \cite{ML:65}.
Considerable further contributions to understanding of generic
properties of the 1D electron liquid have been made in
papers\cite{DzLar:73,EfLar:76,Hal:81}. In particular,
Haldane\cite{Hal:81} has coined the notion of a Luttinger liquid
(LL), stressing the generic properties of the Luttinger model for
1D Fermi systems, and has formulated fundamentals of a modern
bosonization technique as one of the most elegant ways for solving
the problem. In this approach, the Fermi creation and annihilation
operators are explicitly represented in terms of Bose operators
and a 4-fermionic Hamiltonian is eventually diagonalized in the
bosonic representation.

There exists an alternative way to bosonize the problem (sometimes
called 'functional bosonization') which was suggested in
\cite{Fogedby:76} and found its further elaboration in papers
\cite{LeeChen:88,fernandez01-1,Yur}. In this paper, we will apply
a functional method developed earlier by one of us \cite{Yur} for
the treatment of a single-impurity problem in the Luttinger model.
The essence of the method is in the decoupling of the four-fermion
interaction by the standard Hubbard-Stratonovich transformation,
typical for higher-dimensional problems, and a subsequent
elimination of a mixed fermion-boson term in the action by a gauge
transformation which is exact for the pure 1D Luttinger model and
gives a convenient starting point for including a single
backscattering impurity.

The problem of the impurity in the LL is, of course, not new.
However, it was a subject of some controversy. Although it is
generally accepted\cite{furusaki,fabrizio,delft}  that the local
density of states (LDoS)  at the impurity site is suppressed in
agreement with a generic consideration due to Kane and Fisher,
\cite{kane} an alternative prediction of its enhancement as also
been made.\cite{oreg}  Although our method is rather similar to
that of Ref.~\onlinecite{oreg}, we show that its consistent
application leads to the results of
Refs.~\onlinecite{furusaki,fabrizio,delft,kane}. Moreover, we
obtain a general expression for the Green function applicable in a
wide range of temporal and spatial arguments that contains not
only the suppression of the LDoS at the impurity cite, but also
describes its dependence on the distance from the impurity, as
well  as the Friedel oscillations
\cite{fernandez01-1,egger95-1,leclair96-1}. Such a  generality
show the suggested formalism to be quite powerful and useful.

\section{Exact functional representation for Green function}
\setlength{\mathindent}{0cm}

The Hamiltonian of Luttinger liquid with one backscattering impurity can be written as:
\begin{align}\label{H}
\hat{H}=&-i\,a \vf \int\! \dd x \hat{\psi}^{\dagger}_{\eta}(x)
\frac{\partial}{\partial x}\hat{\psi}_{\eta}(x)\nonumber\\
&+\frac{1}{2}\int \! \dd x \dd x
'\hat{\psi}^{\dagger}_{\eta}(x)\hat{\psi}^{\dagger}_{\eta'}(x')V_0(x-x')
\hat{\psi}_{\eta'}(x')\hat{\psi}_{\eta}(x) \nonumber\\
&+ \vf \int\!\dd x\lambda(x) \left[\hat{\psi}^{\dagger}_+(x)\hat{\psi}_-(x)+
\hat{\psi}^{\dagger}_-(x)\hat{\psi}_+(x)\right]
\end{align}
Here $\psi^\dagger_\pm$ and $\psi_\pm$ are the standard creation
and annihilation operators for left- and right-moving electrons
($\psi=\psi_+e^{i\pf x}+\psi_-e^{-i\pf x}$), $\eta=\pm$ and the
summation over repeated indices is implied; $V_0$ is a bare
electron-electron interaction. The impurity potential is given by
$\vf\lambda(x)=\vf\lambda u(x)$, where $u(x)$ is some
 form-factor of the impurity and $\lambda\ll 1$ is its strength.

The temperature Green functions of the Hamiltonian (\ref{H}) can be represented by the functional integral:
\begin{align}
\mathcal{G}_{\eta\eta'}(\xi  ;\xi  ')
=&Z^{-1}_{\lambda}\int\!\psi_{\eta}(\xi ) \psi^*_{\eta'}(\xi
')\,\e^{-S[\psi]}\,\mathcal{D}\psi \label{G}
\end{align}
where $\xi  \equiv (x,\tau )$, $\displaystyle Z_{\lambda}= \int\! {\e}^{-S[\psi]}\,\mathcal{D}\psi$, and the
action
\begin{align}
S[\psi]&= \int\!\dd \xi  \Bigl[
 \psi^*_{\eta}(\xi  )\partial_{\tau}\psi_{\eta}(\xi  )-H(\psi^*,
\psi)\Bigr]\,.
\end{align}
Here the  integration over the imaginary time $\tau$ goes from $0$ to $\beta= 1/T$, and the ``classical''
Hamiltonian $H$ is obtained from $\hat H$, \Eq H,  by substituting the Grassmann fields $\psi^*(\xi  )$ and
$\psi(\xi  ) $ for $\hat\psi^\dagger$ and $\hat\psi$.

The Green function of real electrons is given as a sum of all
$\mathcal{G}_{\eta\eta'}$ taken with appropriate phase factors:

\begin{align}
\mathcal{G}(\xi,\xi')=
\sum\limits_{\eta,\eta'}\mathcal{G}_{\eta\eta'}(\xi,\xi')e^{i\pf(\eta
x-\eta'x')}
\end{align}

The functional bosonization is achieved via the standard
Hubbard-Stratonovich transformation decoupling the 4-fermion term
in the action. Introducing a new classical (bosonic) field $\phi$
we arrive at the action
\begin{align}
S[\phi,\psi] &=
\int\!\dd \xi  \,\psi^*_{\eta}(\xi  )\left(\partial_{\eta}-i\phi\right)\psi_{\eta}(\xi  )\nonumber\\
+\frac{1}{2}&\int \!\dd \xi  \dd \xi  ' \,\phi(\xi  )\, V^{-1}_0(\xi \!-\!\xi  ')\,\phi(\xi  ')
\notag\\\label{S} -\vf &\int \!\dd \xi\, \lambda(x) \Big[\psi^*_+ (\xi  )\psi_-(\xi  ) +\psi^*_-(\xi )
\psi_+(\xi  ) \Big]
\end{align}
Here $V^{-1}_0$ is a function inverse to $V_0$ in the operator
sense, whose $\tau$ dependence is just $\delta ({\tau} \!-\!{\tau}
')$, and the `chiral derivatives' $\partial_{\eta}$ are defined by
\begin{align*}
\partial_+&\equiv 2\partial_z=\partial_{\tau}-i \vf \partial_x\,,&
\partial_-&\equiv 2\partial_{\bar z}=\partial_{\tau}+i \vf \partial_x
\end{align*}

In order to eliminate the mixed term in the chiral derivative in \Eq S, we apply the gauge transformation
 \begin{align}\label{gauge}
\psi_{\eta}&\mapsto\psi_{\eta}e^{i\theta_{\eta}}\,,& \psi^*_{\eta}
&\mapsto\psi^*_{\eta}e^{-i\theta_{\eta}}\,,&\partial_{\eta}\theta_{\eta}&=\phi\,,
\end{align}
where $\theta_{\eta}$ is a complex function which depends on the
field $\phi$. Since $\phi$ is real, $$\theta_-=\theta^*_+\equiv
\theta\equiv \theta_1+i\theta_2\,.$$ This transformation
produces\cite{Yur} the Jacobian $J$, derived explicitly in the
Appendix,
\begin{align}\label{J}
&\ln J=-\frac{1}{2}\int\!\!\dd \xi  \dd \xi  '\, \phi(\xi )\Pi(\xi \!-\!\xi  ')\phi(\xi  ')\,,
\end{align}
where $\Pi$ is the polarization operator whose Fourier transform is given by
\begin{align}\label{P}
\Pi(q,\Omega)=\frac{1}{\pi \vf }\frac{v^2_{\text{\sc f}}q^2}{\Omega^2+v^2_{\text{\sc f}}q^2},
\end{align}
where $\Omega=2\pi nT$ is a bosonic Matsubara frequency,

Therefore, after the transformation \Ref{gauge} the interaction in
the action \Ref S becomes RPA-screened as expected,
\mbox{$V_0\mapsto V^{-1}=V^{-1}_0+\Pi$}, and can be split into the
sum $S[\psi,
\phi]=S_{\text{f}}[\psi]+S_{\text{b}}[\phi]+S_{\text{imp}}[\psi,\phi]$:
\begin{align}
S_{\text{b}}&=\frac{1}{2} \int \!\!\dd \xi  \dd \xi  ' \phi(\xi )\, V^{-1} (\xi  \!-\!\xi  ')\,\phi(\xi
')\notag
\\\label{S3}
S_{\text{f}}&= \int\!\!\dd \xi  \, \psi^*_{\eta}(\xi
)\,\partial_{\eta}\,\psi_{\eta}(\xi  )
\\ S_{\text{imp}}&=-\vf\!\!
 \int\!\!\dd \xi  \lambda\, \Big[\e^{2\theta_2 } \psi^*_+(\xi  ) \psi_-(\xi  )
+ \e^{-2\theta_2 }\psi^*_-(\xi  ) \psi_+ (\xi  )\Big]\notag
\end{align}
The Green function \Ref G can be represented as the functional average over the fermionic and bosonic fields
with the weight $S_{\text{b}}+S_{\text{f}}$:
\begin{align}\label{f01}
\mathcal{G}_{\eta\eta'}(\xi  ;\xi  ')&=\frac{ \cum{
\e^{i\theta_{\eta}(\xi )-i\theta_{\eta'}(\xi  ')} \psi_{\eta}(\xi
) \psi^*_{\eta'}(\xi ')\e^{-S_{\text{imp}}}}} {\cum{
\e^{-S_{\text{imp}}}}}
\\[4pt]
 \cum{\mathcal{O} [\phi,\psi]}&\equiv
 \frac{\int\mathcal{D}\phi\mathcal{D}\psi
\,\mathcal{O}[\phi,\psi]\,\e^{-S_b[\phi]-S_f[\psi]}}{\int\mathcal{D}\phi\mathcal{D}\psi\,
\e^{-S_b[\phi]-S_f[\psi]}}\,.\nonumber
\end{align}
The  bosonic field $\phi$  enters the pre-exponential factor in
\Eq{f01} only implicitly, via $\theta(\phi)$.   Before proceeding
further it is convenient to work out correlation functions of
$\theta$ which follow straightforwardly from
$\bigl<\phi(\xi)\phi(\xi')\bigr>_\phi=V(\xi\!-\!\xi')$ and
\Eq{gauge}:
\begin{align}
\bigl<\theta_1(\xi )\theta_1(\xi ' )\bigr>_\phi =&\frac{1}{2 }\ln\frac{\,|\sin(\zf-\zf')|
^{\phantom{1/g}}}{\, |\sin(z -z ')|^{1/g}}  \nonumber\\
\bigl<\theta_2(\xi )\theta_2(\xi ' )\bigr>_\phi =&\frac{1}{2}\ln \frac{|\sin(z -z ')|^g}
  {|\sin(\zf-\zf')|^{\phantom{g}}} \label{f02}  \\
\bigl<\theta_1(\xi )\theta_2(\xi ' )\bigr>_\phi =&\frac{1}{2}\arg\frac{\sin(z -z ')}{\sin(\zf-\zf')}\,,\nonumber
\end{align}
where
\begin{align}\label{zzf}
\zf&=\pi T(\tau+ix/\vf )   &z& =\pi T(\tau+ix/v)\notag\\
\mbox{\,}\\[-16pt] v&=\vf \left({1+\frac{V(q\!=\!0)}{\pi \vf
}}\right)^{1/2}
 &g&=\frac{\vf }{v} \,.\notag
\end{align}
Here  we assumed  that the Fourier transform of the
forward-scattering pair interaction  only weakly depends on
momentum, i.e.\ ${ V}(q\!\ll \!2\pf)\approx V(q\!=\!0)$.

Now we reduce the partition function $ Z_{\lambda}=\big<\!\big<{\e^{-S_{\text{imp}}}} \big>\!\big>$ in
Eq.~(\ref{f01}) to the product of fermionic and bosonic integrals. This can be done for an arbitrary scattering
potential $\lambda(x)$. On expanding $\e^{-S_{\text{imp}}}$ and keeping only the terms with equal numbers of
$\psi^*_+$ and $\psi_+$ (as well as of $\psi^*_-$ and $\psi_-$), we obtain
\begin{align}
Z_{\lambda}&=\sum^{\infty}_{n=0}\frac{\vf^{2n} }{(n!)^2}\prod ^{n}_{k=1} \int \!\!\dd\xi_k\dd \xi '_k
\lambda(x_k )\lambda(x'_k )\nonumber\\
&\times\bigl<\prod ^{n}_{k=1}\psi^*_+(\xi_k)\psi_-(\xi_k)
\psi^*_-(\xi'_{k})\psi_+(\xi'_{k})\bigr>_{\psi}\label{f03}\\
&\times\bigl< \exp\big[{2\sum\limits^{n}_{k=1}\big(\theta_2(\xi_k)
-\theta_2(\xi'_{k})\big)}\big]\bigr>_{\phi};\nonumber
\end{align}
 Carrying out the bosonic average with the help of formulae (\ref{f02}),
 we find
\begin{equation}
\bigl< \exp\big[{2\sum\limits^{n}_{k=1}\big(\theta_2(\xi_k) -\theta_2(\xi'_{k})\big)}\big]\bigr>_{\phi}=
\frac{\alpha^{2gn}|P_n(z)|^{2g}} {\alpha^{2n}|P_n(\zf)|^2}\,,\label{f04}
\end{equation}
where
\begin{align*}
 \alpha\sim T/\ef \ll 1
\end{align*}
is a cutoff parameter, and $P_n(z)$ is given by
 \begin{align}
P_n(z)=\frac{\prod\limits^{n}_{i<j}\sin(z _i-z _j) \sin(z '_i-z
'_j)}{\prod\limits^{n}_{i,j=1} \sin(z _i-z '_j)};
\end{align}
The parameters $z$ and $\zf$, entering with the appropriate
indices \Eq{f04}, are defined by \Eq{zzf}.

The fermionic average in \Eq{f03} is independent for the left- and
right-moving electrons and yields
\begin{equation}\label{gf}
\bigl<\prod^{n}_{k=1}\psi^{\!\phantom{*}}_-(\xi_k)\psi^*_-(\xi'_k)\bigr>
=\left(\frac{T}{2\vf }\right)^n \!{\rm det}\frac{1}{\sin(\zfi
i-\zfi j')} \,.
\end{equation}
Applying the Cauchy formula \cite{zinnjustin},
\begin{equation*}
{\rm det}\frac{1}{\sin(z_i-z_j')}= (-1)^{\frac{n(n-1)}{2}}\,P_n(z)\,,
\end{equation*}
one sees that the fermionic average \Ref{gf} cancels $|P_n(\zf)|^2$ in the denominator of \Eq{f04}  so that
\begin{align}
Z_{\lambda}=&\sum ^{\infty}_{n=0}\frac{1}{(n!)^2}\left(\frac{T}{2\alpha^{1-g}}\right)^{2n}
 \nonumber\\
&\times\prod ^{n}_{k=1}\int\!\! \dd \xi_k\dd \xi'_k\, \lambda(x_k)\lambda^*(x'_k)\,|P_n(z)|^{2g}\,.\label{f05}
\end{align}
The result above has been obtained by formally calculating both fermionic and bosonic integrals for the
partition function in \Eq{f01}. Now we `re-bosonize' this expression by presenting it as a result of the
integration over a new bosonic field ${\Theta} $:
\begin{align}
\label{Z2} Z_{\lambda}=\left\langle e^{-S_{\lambda}[\Theta]}\right\rangle_0
\end{align}
where
\begin{align}
S_{\lambda}[\Theta]&=-\frac{T}{\alpha}\int \!\!\dd \xi\,\lambda(x ) \cos\Theta(\xi )\label{S1}
\end{align}
and $\langle...\rangle_0$ average is defined with the action $S_0$:
\begin{align}
\label{S2} S_0[\Theta]&=\frac{1}{8\pi g v}\int \dd \xi
\left[(\partial_{\tau}\Theta)^2+v^2(\partial_x\Theta)^2\right]
\end{align}
To verify the validity of the representation \Ref{Z2}--\Ref{S1} one needs to expand the exponent in \Eq{Z2}
using the fact that the pair correlation function of ${\Theta} $ with the action $S_0$ is given (with a proper
regularization) by
\begin{equation}\label{G0}
G_0(\xi,\xi')\equiv \bigl<\Theta(\xi )\Theta(\xi ' )\bigr>_0 =-2g\ln|\sin(z -z ')|\,.
\end{equation}
The sum  resulting from such an expansion coincides with that in Eq.~(\ref{f05}). We remind that here $\xi$
stands as a short hand for $x, \tau$ with the appropriate indices, while $z=\pi T(\tau + ix/v)$, \Eq{zzf}.

On repeating the steps outlined in Eqs.~(\ref{f03})--(\ref{f05}), we obtain the following
$\Theta$-representation for the  Green function of Eq~(\ref{f01}):
\begin{align}
\mathcal{G}_{\eta\eta'}(\xi ,\xi ') =\frac{T}{2\vf
\alpha^{1-1/2g}}\frac{ s_{\eta\eta'}(z \!-\!z ')} {|\sin(z\!-\!z
')|^{\frac{1}{2g}}}\,\widetilde{\mathcal{G}}_{\eta\eta'}(\xi,\xi')\,,
\label{GF}
\end{align}
where we  introduced an auxiliary function
$\widetilde{\mathcal{G}}_{\eta\eta'}(\xi,\xi')$:
\begin{align}
\widetilde{\mathcal{G}}_{\eta\eta'}(\xi,\xi')=Z^{-1}_{\lambda}
\left\langle{\displaystyle\e^ {\frac{ia}{2}\Theta(\xi)
-\frac{ia'}{2}\Theta(\xi ' ) }\e^{
-S_{\lambda}\left[\Theta-\chi\right]
}}\right\rangle_0\,.\label{f12}
\end{align}
The $\cos \Theta$ term in action $S_{\lambda}$, \Eq{S1}, is now shifted by the phase factor  $\chi(\xi _1)$
\begin{align}
    S_{\lambda}[\Theta-\chi]&=-\frac{T}{\alpha}\int \!\!\dd
    \xi_1\,\lambda(x_1
) \cos\left\{\Theta(\xi_1 )-\chi(\xi_1)\right\}\label{S4}
\end{align}
where $\chi(\xi_1)$ parametrically  depends on the arguments of the Green function \Ref{GF},
\begin{align}
 \chi(x_1,\tau_1)\equiv
 \chi(z _1)=\arg\frac{\sin(z _1-z )}{\sin(z _1-z ')}\,. \label{chi}
\end{align}
Finally, the sign factor  $s_{\eta\eta'}$ in \Eq{GF} is defined
(with $\eta=\pm1$) by
\begin{align*}
     s_{\eta\eta'}(z -z ')=
     \exp\left[\tfrac12i(\eta+\eta')\arg\,\sin(z-z')\right]\,.
\end{align*}

In the particular case $x=x'=0$, the representation \Ref{GF}--\Ref{f12} coincides with that obtained by Oreg and
Finkel'stein\cite{oreg}, although we will show below  that it leads to the well known result \cite{kane} for the
density of states rather than that declared in Ref.\onlinecite{oreg}.

\section{Self-consistent harmonic approximation}

The  representation (\ref{GF})--\Ref{f12} is  exact and  further calculations are only possible after some
approximations. First, we assume the impurity to be point-like, $\lambda(x)=\lambda\,\delta(x)$, and weak, i.e.\
$\lambda\ll 1$. This ensures the spectrum linearization to be valid in the presence of the impurity and all the
relevant energy scales to be small compared to the Fermi energy.

It is well known that even such a ``weak'' impurity leads to strong changes to the single-electron density of
states in its vicinity. It's influence is perturbative only in the high-temperature limit,
$\lambda\ll\alpha\propto T/\ef$. In the low-temperature regime, $\lambda\gg\alpha$, a non-perturbative approach
is required. In the present context, it can be developed within the so called self-consistent harmonic
approximation (SCHA) \cite{saito}. It is based on the fact that for $\alpha\ll\lambda\ll 1$, the deviation of
$\Theta$ from $\chi(\tau_1)$ in the action \Ref{f12} is prohibitive so that their difference can be presented as
small quadratic fluctuations around one of the minima of $\cos({\Theta} -\chi)$. By minimizing the difference
between actual cosine-shaped potential and its quadratic fit (the Feynmann-Vernon variational principle), one
substitutes the exact cosine potential by the harmonic one thus reducing the action \Ref{S4} to the following
one:
\begin{equation}\label{subs}
S_\Lambda[{\Theta} -\chi]=\frac{\Lambda T} {2{\alpha} }\int\!\dd{\tau} _1\left[\Theta(0,{\tau} _1)-\chi(0,{\tau}
_1)\right]^2,
\end{equation}
where the new 'impurity strength' $\Lambda$ is chosen to provide the best fit to the real potential. As a
result, one obtains \cite{saito} the renormalized `self-consistent' impurity strength ${\Lambda} $ as
$\Lambda=\lambda^{\frac{1}{1-g}}$.

Now the action is quadratic in $\Theta$,
\begin{align}\label{S5}
S[\Theta]=S_0[\Theta]+\frac{\Lambda T}{2\alpha}\int d\tau\Theta^2(0,\tau)
\end{align}
so the integral \Ref{f12} for
$\widetilde{\mathcal{G}}_{\eta\eta'}(\xi,\xi')$ is reduced to
calculating the  averages with the action \Ref{S5} of linear in
${\Theta} $ terms in the exponent, using the standard formulae of
the type
\begin{align*}
    \av{\e^{b{\Theta} }}_{\Theta} =\exp\left[\frac{b^2}{2}\av{{\Theta} ^2}_{\Theta}
    \right]\,,
\end{align*}
where $\av{\ldots}_{\Theta} $ stand for the functional averaging with the action \Ref{S5}. The integration thus
yields
\begin{align}
-\ln{\widetilde{\cal G}}_{\eta\eta'}(\xi,\xi')
=&\frac{1}{8}G(\xi,\xi)+\frac{1}{8}G(\xi',\xi')-
\frac{\eta\eta'}{4}G(\xi,\xi')\nonumber\\
&-i\Phi_{\eta\eta'}(\xi,\xi')+\Xi(\xi,\xi')\,.\label{f101}
\end{align}
All the terms above can be expressed via the pair correlation function of the auxiliary bosons ${\Theta} $
defined by
\begin{align}\label{G1}
G(\xi,\xi')=\av{\Theta(\xi)\Theta(\xi')}_{\Theta}\,.
\end{align}
In the absence of the ${\Lambda} $ term in action \Ref{S5}, $G$
reduces to the standard bosonic Green  function $G_0$, \Eq{G0}.
The full Green function $G$ in the presence of impurity is
straightforward to find in the Matsubara frequency representation,
\begin{align}
G(\xi;\xi')&=T\sum_{\omega} G(x,x';\omega)\e^{-i{\omega} ({\tau} -{\tau} ')}\,,\notag
 \end{align}
 where it is expressed via $G_0(x,x';\omega)$ as follows:
 \begin{align}\label{G2} G(x,x';\omega)&=G_0(x,x';\omega)-\frac{\Lambda T}{\alpha}
 \frac{G_0(x,0;\omega)G_0(0,x';\omega)}{1+\frac{\Lambda T}{\alpha}
 G_0(0,0;\omega)}\notag\\&=\frac{2\pi g}{|\omega|}
\bigg[e^{-\frac{|\omega|}{v}|x-x'|} - \frac{\e^{-\frac{|\omega|}{v}(|x|+|x'|)}}{\frac{{\alpha} |{\omega} |}{2\pi
g{\Lambda} T}+1 }\bigg]\,.
\end{align}
Exponentiating the denominator of the second term above by $ {1}/{D}=2\int\! \dd s\, \e^{-2Ds}$, we obtain $G$
in the $x,\tau$ representation:
\begin{equation}\label{f92}
G(\xi,\xi')-G_0(\xi,\xi')
=\qquad\qquad\qquad\qquad\qquad\\[-3mm]\end{equation}
\begin{equation}\nonumber =4g\int\limits_{0}^{\infty}\!\!\dd s \e^{-2s}
\ln\!\left|\sin\!\left[\pi T\left(\tau\!-\!\tau'+ i\frac{|x|\!+\!|x'|}{v}\right)+i\frac{\alpha\,s}
{g\Lambda}\right] \right|
\end{equation}

  The impurity-induced terms in \Eq{f101}, $\Phi$ and $\Xi$, which
are  respectively   linear and quadratic in the factor $\chi$, \Eq{chi}, result from the averaging of the first
and zeroth order terms in ${\Theta} $  arising from \Eq{subs}.
\begin{align}\label{Xi}
\Xi&=\frac{\Lambda T}{2\alpha}\int{\rm d}\tau_1{\rm d}\tau_2\Big[ \frac{\Lambda
T}{\alpha}G(0,\tau_1;0,\tau_2)\notag\\& \quad-\delta(\tau_1-\tau_2)\Big] \chi(\tau_1 )\chi(\tau_2 )
\\\label{Phi}
\Phi_{\eta\eta'} &=\frac{\Lambda T}{2\alpha}\int{\rm
d}\tau_1\left[ a\,G(\xi;0,\tau_1)-a'\,G(\xi';0,\tau_1)\right]
\chi(\tau_1 )
\end{align}
It should be stressed that both $\Xi$ and $\Phi$ depend on  the
`observation points' $\xi$ and $\xi'$ via the appropriate
dependence of the parameter $\chi$, \Eq{chi}.  All these functions
can be calculated for arbitrary $\xi$ and $\xi'$. However,  since
we are only interested in the local density of states (at an
arbitrary distance from the impurity) and Friedel oscillations, it
is sufficient to consider  $x=x'$ case only; $\tau'$ for
convenience is set to zero so that from now on we  use
$\xi=(x,\tau) $ and $\xi'=(x,0)$. Then we obtain
\begin{align}\label{xi}
\Xi(x,\tau)=\,&\frac{\Lambda
T^2}{2\alpha}\sum\limits_{\omega}\left|
\chi(0,\omega)\right|^2\left[1-\frac{\Lambda T}{\alpha}
G(0,0;\omega)\right]\\\label{phi}
\Phi_{\eta\eta'}(x,\tau)&=\frac{\Lambda
T^2}{2\alpha}\sum\limits_{\omega}\chi(0,\omega)
\big[aG(x,0;\omega)e^{-i\omega\tau}-\nonumber\\
&-a'G(0,x;\omega)\big].
\end{align}

Substituting the Fourier transform of \Eq{chi},
\begin{align}\label{chi2}
\chi(0,\omega)=\frac{i\pi}{\omega}\sgn\,x\,e^{-|\omega|\frac{|x|}{v}}\Big[e^{i\omega\tau}-1\Big]\,,
\end{align}
and \Eq{G2} into Eqs.~\Ref{xi} and \Ref{phi},  we carry out the
Matsubara summation to obtain $\Xi$ and $\Phi$ in the same
representation as follows:
\begin{align}
&\Xi(x;\tau)=\frac{1}{2g}\int\limits_{0}^{\infty}\!\!\dd s\,\e^{-2s} \ln\left[1+\frac{\sin^2\pi T\tau}{\sinh^2\!
\left(\frac{\alpha\,s}{g\Lambda}+2\pi T\frac{|x|}{v}\right)}\right] \notag\\
\label{f38}\\
&\Phi_{\eta\eta'}(x;\tau)=\frac{\eta\!-\!\eta'}2\sgn x\biggl[
\frac{\pi}{2}- \Im \text m\ln\sin\pi T\left(\tau+2i\frac{|x|}{v}
\right)\biggr]\notag
\end{align}

Now we have all the ingredients to find the Green function
\Ref{GF}. Using \Eq{f101}, we express $ {\mathcal G}(\xi)$   in
terms of $G$, $\Phi$, and $\Xi$ as follows:
\begin{align}
&\mathcal{G}(\xi )= \frac{T}{\vf }\frac{\alpha^{\frac{1}{2g}-1}}{|\sin\pi T\tau|^{\frac{1}{2g}}}
\,\e^{-\Xi(\xi )}\e^{-\frac{1}{4}G(x,x;\tau=0)}\notag \\
&\times\left[\sgn\tau\,\e^{\frac{1}{4}G(\xi)}-\e^{-\frac{1}{4}G(\xi)} \cos[2\pf
x+\Phi_{+-}(\xi)]\right]\label{f20}
\end{align}
The appropriate asymptotic behavior of $G$, $\Phi$, and $\Xi$ is
determined from formulae \Ref{f92},   (\ref{f38}). In region
$g|x|\pf \ll\Lambda^{-1},\,\,|\tau|\ef\ll\Lambda^{-1}$ the
impurity can be described perturbatively and its presence results
only in a small correction to the Green function of the pure LL.
So we restrict our attention to $G(\xi)=\mathcal{G}(x,x;\tau)$
outside of the perturbative region on the $(x,\tau)$ plane.
Introducing convenient dimensionless notation
$\tilde{x}=g\pf|x|,\,\, \tilde{\tau}=\ef|\tau|$ and keeping only
the leading contributions, we find
\begin{widetext}\begin{align*}
G(x,x;\tau) =\left\{
\begin{array}{cl}
\displaystyle 2g\ln\Lambda^{-1}, & \tilde{x}\ll\Lambda^{-1};\,\tau=0\\
&\\
\displaystyle 2g\ln\frac{\sinh\alpha\tilde{x}}{\sin\alpha\tilde{\tau}}, &
\tilde{x}\gg\Lambda^{-1},\tilde{\tau}\\
&\\
\displaystyle 0, & \tilde{\tau}\gg\Lambda^{-1},\tilde{x}\\
\end{array}\right.\qquad
\Xi(x,\tau)=\left\{
\begin{array}{cl}
\displaystyle \frac{1}{2g}\ln\frac{\sin\alpha\tilde{\tau}}{\alpha\tilde{x}},
& \Lambda^{-1}\ll\tilde{x}\ll\tilde{\tau}\\
&\\
\displaystyle \frac{1}{2g}\ln\frac{\Lambda\sin\alpha\tilde{\tau}}{\alpha},
& \tilde{x}\ll\Lambda^{-1}\ll\tilde{\tau}\\
\phantom{\frac{1}{2g}\ln\frac{\sin\alpha\tilde{\tau}}{\alpha\tilde{x}} }&\\
\displaystyle 0, & \tilde{x}\gg\Lambda^{-1},\tilde{\tau}
\end{array}\right.\\[8pt]
\end{align*}
We do not write asymptotics for $\Phi_{+-}(x,\tau)$ explicitly
since the phase does not make any real impact. The above
expressions enable us to find the Green function, \Eq{f20}, for
any distance $x$ from the impurity:

\begin{align*}
\mathcal{G}(x,x;\tau) & =\left\{
\begin{array}{cll}
&\displaystyle \frac{T}{\vf} \frac{\alpha^{\frac{1}{g}-1}} {(\sin\alpha\tilde{\tau})^{\frac{1}{g}}}
\left(\Lambda^{-1}\right)^ {\frac{1}{2}\left(\frac{1}{g}-g\right)} \Big[\sgn\,\tau-\cos (2\pf x+\Phi_{+-})\Big],
& \,\,\,\tilde{x}\ll\Lambda^{-1}\ll\tilde{\tau}\\
& \phantom{0} & \phantom{0} \\
&\displaystyle \frac{T}{\vf} \frac{\alpha^{\frac{1}{g}-1}} {(\sin\alpha\tilde{\tau})^{\frac{1}{g}}}\,
\tilde{x}^{\frac{1}{2}\left(\frac{1}{g}-g\right)} \Big[\sgn\,\tau-\cos (2\pf x+\Phi_{+-})\Big], & \,\,\,
\Lambda^{-1}\ll\tilde{x}\ll\tilde{\tau}\\
& \phantom{0} & \phantom{0} \\
& \displaystyle \frac{T}{\vf } \frac{\alpha^{\frac{1}{2}\left(\frac{1}{g}+g\right)-1}}
{(\sin\alpha\tilde{\tau})^{\frac{1}{2} \left(\frac{1}{g}+g\right)}}
\Big[\sgn\,\tau-\left(\frac{\sin\alpha\tilde{\tau}} {\sinh\alpha\tilde{x}}\right)^g\cos (2\pf x+\Phi_{+-})\Big],
& \,\,\,\tilde{x}\gg\Lambda^{-1},\tilde{\tau}
\end{array}
\right.
\end{align*}
\end{widetext}

To begin with, the general expression for the Green function,
\Eq{f20}, allows us  to extract a well known result for the
Friedel oscillations: \cite{fernandez01-1,egger95-1,leclair96-1}
\begin{align}
\Delta\rho\left(x\gg (\Lambda \pf)^{-1}\right)=\frac{T}{\vf \alpha^{1-g}}\frac{1}{\left|\sinh\frac{2\pi
Tx}{v}\right|^g}\cos 2\pf x;
\end{align}
For $x$  in the range $(\pf\Lambda)^{-1}\ll\, x\, \ll \ell_T$
($\ell_T$ being  the thermal dephasing length) the amplitude of
the Friedel oscillations decreases as $\sim |x|^{-g}$, and for
larger $x$ it is exponentially suppressed at distances exceeding
$\ell_T/2g$.

More interesting is the  local density of states (LDoS) which is defined via an appropriate analytical
continuation of the Fourier transform of $\mathcal{G}$:
\begin{align*}
\nu(x,\varepsilon)&=-\frac{1}{\pi}\left.\Im{\text{m}}\int \dd \tau\,e^{-i\epsilon\tau}\,\mathcal{G}(\xi
)\right|_{i\epsilon=\varepsilon}
\end{align*}
Using asymptotics for the Green function, we get:
\begin{align}
\nu(x,\varepsilon)\sim\left\{
\begin{array}{cll}
&\displaystyle \varepsilon^{\frac{1}{g}-1} \Lambda^{-\frac{1}{2}\left(\frac{1}{g}-g\right)},
&\,\,\, \tilde{x}\ll\Lambda^{-1}\ll\varepsilon^{-1} \\
& \phantom{0} & \phantom{0} \\
& \displaystyle \varepsilon^{\frac{1}{g}-1}\tilde{x}^{\frac{1}{2}
\left(\frac{1}{g}-g\right)}, &\,\,\, \Lambda^{-1}\ll\tilde{x}\ll\varepsilon^{-1} \\
& \phantom{0} & \phantom{0} \\
& \displaystyle \varepsilon^{\frac{1}{2}\left(\frac{1}{g}+g\right)-1},
 & \,\,\, \tilde{x}\gg\Lambda^{-1},\varepsilon^{-1}
\end{array}
\right. \label{ldos}
\end{align}
The three  regions above with different behavior of LDoS are sketched in Fig.1.  The first line in (\ref{ldos})
describes LDoS in the vicinity of impurity, in full correspondence with the original results of Kane and Fisher
\cite{kane} obtained for the LDoS at $x=0$, i.e.\ exactly at the impurity. In addition, we specify here a
dependence of the LDoS on the strength of the impurity (we want to remind that
$\Lambda=\lambda^{\frac{1}{1-g}}$, which leads to LDoS being proportional to $\lambda^{-\frac{1+g}{2g}}$). The
region of applicability of this result corresponds to the diagonally hatched section in Fig.~1.

\begin{figure}[b]
\begin{picture}(10,7)\thicklines
\put(1,1){\vector(1,0){7}} \put(1,1){\vector(0,1){5.5}} \put(2.5,0.9){\line(0,1){0.2}}
\put(6.0,0.9){\line(0,1){0.2}} \put(0.9,2.0){\line(1,0){0.2}} \put(0.9,4.33){\line(1,0){0.2}}
\put(2.4,0.5){$\Lambda^{-1}$} \put(5.9,0.5){$\alpha^{-1}$} \put(7.5,0.5){$\tilde{x}$} \put(0.4,1.9){$\alpha$}
\put(0.4,4.18){$\Lambda$} \put(0.6,5.9){$\tilde{\varepsilon}$}
\put(2.5,4.33){.} \put(2.52,4.29){.} \put(2.535,4.257){.} \put(2.55,4.23){.} \put(2.565,4.195){.}
\put(2.58,4.165){.} \put(2.595,4.135){.} \put(2.61,4.106){.} \put(2.625,4.077){.} \put(2.64,4.049){.}
\put(2.655,4.021){.} \put(2.67,3.994){.} \put(2.685,3.967){.} \put(2.7,3.94){.} \put(2.715,3.915){.}
\put(2.73,3.89){.} \put(2.745,3.865){.} \put(2.76,3.841){.} \put(2.775,3.817){.} \put(2.79,3.793){.}
\put(2.805,3.77){.} \put(2.825,3.74){.} \put(2.845,3.71){.} \put(2.865,3.68){.} \put(2.885,3.652){.}
\put(2.905,3.625){.} \put(2.925,3.597){.} \put(2.945,3.571){.} \put(2.965,3.545){.} \put(2.985,3.519){.}
\put(3.005,3.494){.} \put(3.025,3.469){.} \put(3.045,3.445){.} \put(3.065,3.421){.} \put(3.085,3.398){.}
\put(3.105,3.375){.} \put(3.125,3.353){.} \put(3.145,3.331){.} \put(3.165,3.309){.} \put(3.185,3.288){.}
\put(3.21,3.262){.} \put(3.235,3.237){.} \put(3.26,3.212){.} \put(3.285,3.188){.} \put(3.315,3.16){.}
\put(3.345,3.132){.} \put(3.375,3.105){.} \put(3.405,3.079){.} \put(3.435,3.053){.} \put(3.465,3.028){.}
\put(3.495,3.004){.} \put(3.525,2.98){.} \put(3.555,2.957){.} \put(3.585,2.934){.} \put(3.615,2.912){.}
\put(3.645,2.89){.} \put(3.675,2.869){.} \put(3.705,2.848){.} \put(3.735,2.828){.} \put(3.765,2.808){.}
\put(3.795,2.789){.} \put(3.825,2.77){.} \put(3.855,2.751){.} \put(3.885,2.733){.} \put(3.915,2.715){.}
\put(3.945,2.698){.} \put(3.975,2.68){.} \put(4.005,2.664){.} \put(4.035,2.647){.} \put(4.065,2.631){.}
\put(4.095,2.616){.} \put(4.125,2.6){.} \put(4.155,2.585){.} \put(4.185,2.57){.} \put(4.215,2.555){.}
\put(4.245,2.541){.} \put(4.275,2.527){.} \put(4.305,2.513){.} \put(4.335,2.499){.} \put(4.365,2.486){.}
\put(4.395,2.473){.} \put(4.425,2.46){.} \put(4.455,2.447){.} \put(4.485,2.435){.} \put(4.515,2.422){.}
\put(4.545,2.41){.} \put(4.575,2.399){.} \put(4.605,2.387){.} \put(4.635,2.376){.} \put(4.67,2.362){.}
\put(4.705,2.35){.} \put(4.74,2.337){.} \put(4.775,2.325){.} \put(4.81,2.312){.} \put(4.845,2.3){.}
\put(4.88,2.289){.} \put(4.915,2.277){.} \put(4.95,2.266){.} \put(4.99,2.253){.} \put(5.03,2.241){.}
\put(5.07,2.229){.} \put(5.11,2.217){.} \put(5.15,2.205){.} \put(5.19,2.193){.} \put(5.23,2.182){.}
\put(5.27,2.171){.} \put(5.31,2.16){.} \put(5.35,2.149){.} \put(5.39,2.139){.} \put(5.43,2.129){.}
\put(5.47,2.119){.} \put(5.51,2.109){.} \put(5.55,2.099){.} \put(5.59,2.089){.} \put(5.63,2.08){.}
\put(5.67,2.071){.} \put(5.71,2.062){.} \put(5.75,2.053){.} \put(5.79,2.044){.} \put(5.83,2.035){.}
\put(5.87,2.027){.} \put(5.91,2.018){.} \put(5.95,2.01){.} \put(5.99,2.002){.}

\put(1.0,2.0){\line(1,0){6.5}} \put(1.0,4.33){\line(1,0){1.5}} \put(2.5,2){\line(0,1){2.3}} \thinlines
\put(2.12,2){\line(1,1){0.38}} \put(1.74,2){\line(1,1){0.76}} \put(1.36,2){\line(1,1){1.14}}
\put(1,2.02){\line(1,1){1.5}} \put(1,2.4){\line(1,1){1.5}} \put(1,2.78){\line(1,1){1.5}}
\put(1,3.16){\line(1,1){1.17}} \put(1,3.54){\line(1,1){0.79}} \put(1,3.92){\line(1,1){0.41}}
\put(2.7,4.02){\line(0,-1){2.02}} \put(2.9,3.7){\line(0,-1){1.7}} \put(3.1,3.43){\line(0,-1){1.43}}
\put(3.3,3.22){\line(0,-1){1.22}} \put(3.5,3.04){\line(0,-1){1.04}} \put(3.7,2.89){\line(0,-1){0.89}}
\put(3.9,2.75){\line(0,-1){0.75}} \put(4.1,2.66){\line(0,-1){0.66}} \put(4.3,2.56){\line(0,-1){0.56}}
\put(4.5,2.48){\line(0,-1){0.48}} \put(4.7,2.4){\line(0,-1){0.4}} \put(4.9,2.33){\line(0,-1){0.33}}
\put(5.1,2.27){\line(0,-1){0.27}} \put(5.3,2.21){\line(0,-1){0.21}} \put(5.5,2.16){\line(0,-1){0.16}}
\put(5.7,2.1){\line(0,-1){0.1}} \put(5.85,2.07){\line(0,-1){0.07}} \put(2.63,4.13){\line(1,0){4.87}}
\put(2.75,3.93){\line(1,0){4.75}} \put(2.87,3.73){\line(1,0){4.63}} \put(3.03,3.53){\line(1,0){4.47}}
\put(3.21,3.33){\line(1,0){4.29}} \put(3.38,3.13){\line(1,0){4.12}} \put(3.66,2.93){\line(1,0){3.84}}
\put(3.98,2.73){\line(1,0){3.52}} \put(4.38,2.53){\line(1,0){3.12}} \put(4.91,2.33){\line(1,0){2.59}}
\put(5.57,2.13){\line(1,0){1.93}} \multiput(1.0,4.33)(0,0.2){7}{\line(1,0){6.5}} \put(3.8,3.7){\textbf{BULK}}
\put(1.02,2.9){\textbf{Impurity}} \put(2.65,2.25){\textbf{Crossover}}
\end{picture}
\caption{regions with different behavior of local density of states $\nu(x,\varepsilon)$;  $\tilde{x}\equiv g\pf
|x|,\,\, \tilde{\tau}=\ef|\tau|$ }
\end{figure}
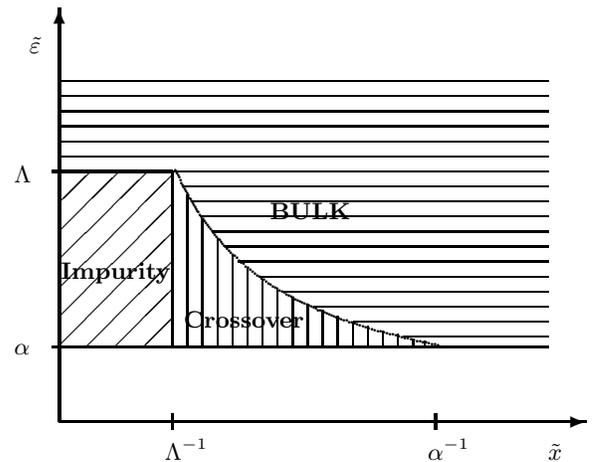

The analytic dependence of the LDoS on the distance from the impurity given by the second line in \Eq{ldos}
describes a new result for the crossover from the impurity-induced dip in the LDoS to the bulk behavior. So, as
a function of ${x}$ at a fixed ${\varepsilon}$, $\nu(x)$ remains constant for $\tilde{x}\lesssim\Lambda^{-1}$,
then it increases until $\tilde{x}$ becomes of order  $\tilde{\varepsilon}^{-1}$ (which happens before $x$
overtakes the thermal dephasing length, $\ell_T=\vf /\pi T$), where the LDoS reaches its bulk ($x$-independent)
value, given by the  last last line in (\ref{ldos}). This corresponds to a well-known result (see
Ref.~\onlinecite{voit} for reviews) for the LDoS in the homogenous Luttinger liquid. Its region of applicability
is horizontally hatched in Fig.~1 (although Eq.~(\ref{ldos}) does not describe the perturbative region
$\tilde{x},\tilde{\varepsilon}^{-1}\ll\Lambda^{-1}$, the presence of the impurity results there only in small
corrections to bulk behavior). The crossover between the impurity-dominated and bulk values is governed by a
universal power law: $\nu(x)\sim\tilde{x}^{\frac{1}{2}\left(\frac{1}{g}-g\right)}.$

In conclusion, we have demonstrated how powerful is the method of ``functional bosonization'' developed here for
the description of a single back-scattering impurity in the Luttinger liquid. It allows one to  get  within a
single formalism the results both for the local density of states at an arbitrary distance from the impurity and
for the Friedel oscillations.

\begin{acknowledgements} We  gratefully acknowledge support
 by the EPSRC
grants GR/R95432 and GR/S29386.
\end{acknowledgements}

\begin{appendix}
\section{Jacobian}

The Jacobian of the gauge transformation Eq.(\ref{gauge}) can be written as
\begin{align*}
\ln\,J[\phi]=\sum_{\eta=\pm}{\rm
Tr}\ln\left|\frac{\partial_{\eta}\!-\!i\phi}{\partial
_{\eta}}\right|
=-\sum_{\eta=\pm}\sum_{n=1}^{\infty}\frac{1}{n}{\rm Tr}\left(
i\phi g_{\eta}\right)^n
\end{align*}
where the Green function of non-interacting right- or left-moving
electrons, obeying $\partial_{\eta}g_{\eta}=1$, are given  by
\begin{equation}\label{gfl}
    g_-(\xi,\xi')=g_+^*(\xi)=\frac T{2\vf}\,\frac1{\sin(\zf-\zf')}
\end{equation}
where $\zf$ is given by \Eq{zzf}.
 The $n$-th order term in $\phi$ is the sum of two
vertices made of the loops $\Gamma_n^{+}$ and  $\Gamma_n^{-}$ with
$n$ external lines corresponding to $\phi$'s, each loop being
built of the $n$ Green  functions $g_\pm$, respectively:
\begin{align}
 {\rm Tr}\left(g_{\eta}\,\phi\right)^n=\int \prod_{k=1}^{n}
&{\dd}x_k{\dd}{\tau}_k\; \Gamma_n^{(a)}(\zfi{_{\!1}};...;\zfi{n})\,\prod_{i=1}^{n}\phi(x_i,\tau_i),
\notag\\[-3pt]\label{I}\\[-6pt]\notag
\Gamma_n^{(a)}(\zfi{_{\!1}};...;\zfi{n})&=\prod_{i=1}^{n}g_{\eta}(\zfi{i}-\zfi{{i+1}})\,,
\end{align}
with the boundary condition $\zfi{n+1}=\zfi{_{\!1}}$. Substituting
$g_{\eta}$ from \Eq{gfl}, one finds
\begin{equation*}
\Gamma_n^+(\zfi{_{\!1}};...;\zfi{n})\propto \prod_{i=1}^{n}\frac{s_i}{s_i-s_{i+1}}, \quad s_i=e^{2i\zfi{i}}\,.
\end{equation*}
One can rewrite  the symmetric part of this vertex, which contributes into the integral in \Eq I, as follows:
\begin{equation*}
\Gamma_n^+(\zfi{_{\!1}};...;\zfi{n})\propto \frac{{\cal
A}_n(s_1,...s_n)}{\prod_{i<j}^{n}(s_i-s_j)}\,\prod_{k=1}^{n}s_k\,.
\end{equation*}
where ${\cal A}_n$ is an absolutely anti-symmetric polynomial built on $n$ variables $s_i$. By power counting,
its order  should be $n(n-3)/2$.  On the other hand,  the minimal possible order of an  absolutely
anti-symmetric polynomial of $n$ variables is $n(n+1)/2$, as follows from the fact that the powers of different
variables should be different for any monomial in order the entire polynomial to be  anti-symmetric. The two
inequalities can only be satisfied for $n \leq 2$ so that ${\cal A}_{n>2}=0$. Therefore, all loops containing
more than two external lines are zero. (Such an  observation was first made by Dzyaloshinskii and Larkin
\cite{DzLar:73} within diagrammatic techniques).

Therefore, we are left with the contributions from the loops with
one or two external lines. The loop with one external line is
proportional to the zero-momentum mode of the Coulomb interaction
and is cancelled, as always, due to electroneutrality. The loop
with two external lines is just the standard polarization operator
in the random-phase approximation (exact for the LL), given in
$(q,\Omega)$-representation by \Eq P so that the Jacobian is
reduced to that in \Eq J in the main text.

\end{appendix}


\end{document}